\documentclass[a4paper]{article}
 \usepackage{psfrag}
\usepackage{graphicx}
\usepackage{amsmath}
\usepackage{hyperref}
\usepackage{color}

\begin{document}

\begin{center}

{\bf \Large Line graphs as social networks }\\[5mm]

{\large M. J. Krawczyk$^{1a}$, L. Muchnik$^{2b}$, A. Ma\'nka-Kraso\'n$^{1c}$ \\and K. Ku{\l}akowski$^{1d}$ }\\[3mm]

{\em

$^1$Faculty of Physics and Applied Computer Science, AGH University of Science and Technology, al. Mickiewicza 30, PL-30059 Krak\'ow, Poland\\
$^2$Information, Operations and Management Systems Department, Stern School of Business at the New York University, 44 West Fourth Street New York, NY 10012, USA

}

{\tt $^a$gos@fatcat.ftj.agh.edu.pl, $^b$levmuchnik@gmail.com, $^c$impresja@gmail.com, $^d$kulakowski@novell.ftj.agh.edu.pl}

\today

\end{center}

\begin{abstract}
The line graphs are clustered and assortative. They share these topological features with some social networks. We argue that this similarity reveals the cliquey character of the social networks. In the model proposed here, a social network is the line graph of an initial network of families, communities, interest groups, school classes and small companies. These groups play the role of nodes, and individuals are represented by links between these nodes. The picture is supported by the data on the LiveJournal network of about $8\times10^6$ people. In particular, sharp maxima of the observed data of the degree dependence of the clustering coefficient $C(k)$ are associated with cliques in the social network.
\end{abstract}

\noindent

{\em PACS numbers:} 

\noindent


\section{Introduction}

In mathematically oriented sociology, a social network is a paradigm. The idea is to express social relations between individuals by weighted or unweigted links between nodes of a graph. Rough as it is, this approximative representation got a wide interest of social scientists \cite{gra,bk1,bk2,fre,insna}. In interdisciplinary areas, the research on networks was boosted by the seminal paper of Watts and Strogatz in 1998 \cite{wat}. Since then, several books are published on various applications of networks \cite{lkd,dog,hand,pasat,new}. Our aim here is to foster a new application of the network formalism: we propose that the structure of some social networks makes them similar to the structure of the line graph, constructed on a scale-free growing network. The line graphs are known for at least 80 years \cite{whi,evla}, but in the above mentioned interdisciplinary stream their relevance seems underestimated. Some recent applications of this kind of graphs can be found in \cite{evla,mokre,lij,nach1,nach2,chi,chi2}.\\

Our argumentation can be sketched as follows. Suppose that the actual structure of some society is a set of cliques or almost fully connected clusters 
which can be identified with families, groups of friends, small companies or interest groups. Each such clique can be represented as a node of an otherwise uncorrelated network. As it was indicated in a recent analysis of information spreading, two small and strongly connected groups are linked by individuals who belong to both of them and contribute in the information transfer between them \cite{apo}. Once we are interested in a construction of a conventional social network, where humans play the role of nodes, the network should be constructed as the line graph of the initial network of families, school classes etc.\\

A fact rarely considered by social network researchers is that social ties represent particular social context such as common interest, association with the same group or collocation. Considering this, the natural organization of humans in nearly fully connected groups linked by their simultaneous  participation in few of them leads to emergence of social graph observed by researchers. In this manuscript we show that a line graph transformation applied to the underlined network of such groups may indeed
lead to a graph with properties commonly observed in social networks.\\

In the next section we summarize the arguments of Newman and Park, that social networks are both transitive (clusterized) and assortative \cite{new2}. In Section 3 we refer to our recent calculations on the line graphs, where we have shown that these graphs are both clusterized and assortative \cite{amk1,amk2}. In Section 4 we describe new data on the network of users of LiveJournal, which also support the above characteristics of social networks. In the same section the data plot is shown on the clustering coefficient $C$ as dependent on the node degree $k$. In Section 5 we compare the plot $C(k)$ with the result of simulations on the line graph, formed from an uncorrelated scale-free network. Last section is devoted to conclusions.\\

\section{Social networks}

In \cite{new2} and literature therein (\cite{n1,n2,n3,gui,tyl} among others) Newman and Park bring examples of social networks which are clusterized and assortative. Let us recall that in unweighted networks, the clustering is measured by the clustering coefficient $C$ defined as

\begin{equation}
C=\sum_i\frac{2y_i}{k_i(k_i-1)}
\end{equation}
where $y_i$ is the actual number of links between neighbours of $i$-th node, and $k_i$ is the degree of this node, i.e. the number of its neighbours.
In other words, the clustering coefficient is the probability that two neighbours of a node are mutually connected. Once a node has zero or one neighbour only, its contribution to $C$ is zero. By clusterized we mean that the clustering coefficient $C$ is clearly larger, than for a random network.
The examples are: the network of film-actor collaborations ($C$=0.20), the collaboration network of mathematicians ($C$=0.15), the network of company directors ($C$=0.59) and an e-mail network ($C$=0.17). In all these examples, the respective values of $C$ for non-clusterized counterparts are smaller by at least one order of magnitude. \\

The assortativity is a tendency of highly connected nodes to have highly connected neighbours. It can be measured by the Pearson correlation coefficient $r$ between degrees of neighboring nodes. Once $r$ is positive for some network, we can classify this network as assortative. Indeed, as demonstrated by Newman in \cite{n2}, the coefficient $r$ is positive for the network of film-actor collaborations ($r$=0.208), the coauthorship networks of mathematicians ($r$=0.12), physicists ($r$=0.363) and biologists ($r$=0.127), the network of company directors ($r$=0.276) and an e-mail network ($r$=0.17). An exception was found for the network of romantic (not necessary sexual) relationships between students at a US high school \cite{rom}, where $r$=-0.029. However, in this particular case we can understand that individuals are not willing to involve third part into their romantic relations. An alternative way to check if a network is assortative or not is to plot the mean degree $k'$ 
 of neighbours of nodes of degree $k$ as dependent on $k$. If the plot $k'(k)$ is ascending, the investigated network is assortative: more connected nodes have on average more connected neighbours. As we checked in \cite{amk2}, for artificially generated uncorrelated scale-free networks $k'$ does not depend on $k$. \\

A widely cited explanation for  emergence of assortativeness and high clustering in social network \cite{new2} is that individuals belong to many groups, and their social connections are limited to members of the groups they belong to. In other words, the social structure is equivalent to a bipartite network of groups and individuals. In this network, individuals are connected to groups they belong to. What is usually observed is a projection of this structure onto just the individuals.
As an outcome of this projection, we have a social network where individuals are connected with probability $p$ if they belong to the same group; otherwise they are not connected. This model network is found to be both clusterized and assortative \cite{new2}.\\

We suggest that applying the line graph transformation to a network of groups such as families, school classes and common interests associations representing cliques of individuals and connected by the same person that simultaneously belongs to several groups, can result in a social network closely resembling the ones we actually observe.\\

This quantitative explanation of the clusterized and assortative structure of social networks agrees with the well-established opinion of social scientists that cohesive small groups are the main motif in a society. In \cite{bk2}, a list is given of four general properties of these groups. These are: the mutuality of ties, the closeness or reachability of group members, the frequency of ties among members and the relatively smaller frequency of the ties among non-members. In the network formalism, all these properties find their formal shape. On the other hand, the quantitative search of communities initiated in \cite{n3} developed into a large branch of science of networks \cite{new}; recent review on this search can be found in \cite{for}.

\section{Line graphs}

From a given graph $G$ of $N$ nodes and $L$ links, a line graph $G'$ can be constructed as follows \cite{bal}. A node of $G'$ is assigned to each link of $G$. Two nodes of $G'$ are linked if and only if the respective links in $G$ shared a node. In this way, the number $N'$ of nodes in $G'$ is equal to the number $L$ of links in $G$. The number $L'$ of links in $G'$ depends on the degree distribution $P(k)$ in $G$. We have shown numerically in \cite{amk1}, that for three kinds of networks (Erd\"os-Renyi networks, the growing exponential networks and the growing scale-free networks) the degree distribution of $G'$ is close to the degree distribution of $G$. Basically, a node of degree $k$ is converted to a clique (fully connected graph) of $k$ nodes and $k(k-1)/2$ links. Further, a link in $G$ joining nodes of degrees $k_1$ and $k_2$ is converted into a node in $G'$ of degree $k_1+k_2-2$ which belongs to two cliques, one of $k_1$ nodes and another of $k_2$ nodes. \\

For an uncorrelated graph $G$ of the mean degree $<k>$ much smaller than $N$ we can assume, that two different neighbours of a node are not mutually linked. A contribution to the clustering coefficient $C$ of a node in $G'$ contains then merely the contributions from two separate cliques. The number of links between $k_1+k_2-2$ neighbours is then $(k_1-1)(k_1-2)/2+(k_2-1)(k_2-2)/2$. For the degree distribution $P(k)$, the clustering coefficient $C$ is

\begin{equation}
C=\frac{\sum_{k_1,k_2}k_1P(k_1)k_2P(k_2)\frac{(k_1-1)(k_1-2)+(k_2-1)(k_2-2)}{(k_1+k_2-2)(k_1+k_2-3)}}{\sum_{k_1,k_2}k_1P(k_1)k_2P(k_2)}
\end{equation}
This formula was used in \cite{amk1} and the results were compared with numerical calculations. Both methods confirm that for $<k>$ greater than 5, the clustering coefficient $C$ is not smaller than 0.5.\\

The same methods were applied to demonstrate that the line graphs constructed on uncorrelated networks are assortative \cite{amk2}. This is a direct consequence of the fact that the neighboring nodes in the line graphs are formed from links sharing a common node in the initial graph. The degree of this common node contributes to the degree of both neighboring nodes in the line graph.

\section{LiveJournal}

LiveJournal \cite{1} is a remarkably popular platform for personal blog management, populated with over 8 million blogs and over 1 million of communities. LiveJournal was among the first of such platforms available online and it still remains one of the most active and popular. Its users manage personal blogs where they share their daily experiences, political views or discuss news events. Users can also comment on posts of other users. \\

Unlike more dynamic systems like Facebook and Twitter that gained their popularity rather recently, LiveJournal is not based on personal messages or applications. Typical LiveJournal post may contain a significant amount of text with embedded images or video and may be followed by discussion that in times exceed thousands of comments. \\

 \begin{figure}[ht]
 %\vspace{0.3cm}
 \centering
 {\centering \resizebox*{12cm}{9cm}{\rotatebox{00}{\includegraphics{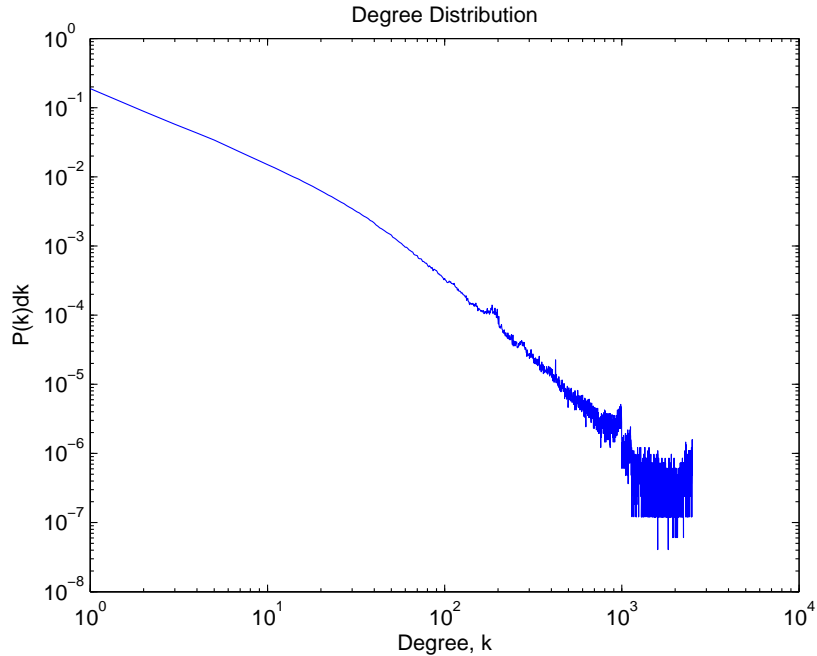}}}}
 %\vspace{0.3cm}
\caption{The degree distribution of the social network of LiveJournal. }
 \label{fig-1}
 \end{figure}

 \begin{figure}[ht]
 %\vspace{0.3cm}
 \centering
 {\centering \resizebox*{12cm}{9cm}{\rotatebox{00}{\includegraphics{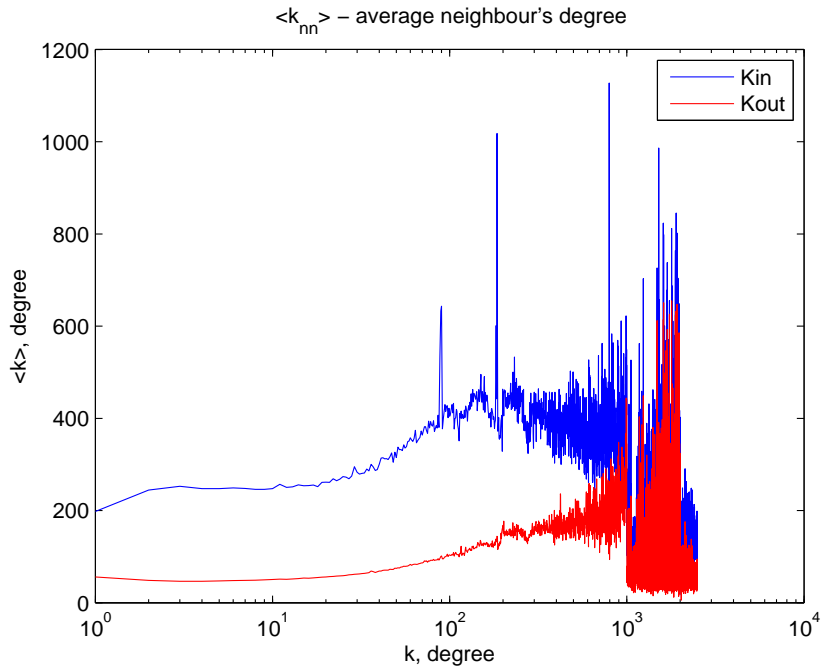}}}}
 %\vspace{0.3cm}
\caption{Mean in-degree (blue) and out-degree (red) of neighbours of nodes of degree $k$ for LiveJournal.}
 \label{fig-2}
 \end{figure}

 \begin{figure}[ht]
 %\vspace{0.3cm}
 \centering
 {\centering \resizebox*{12cm}{9cm}{\rotatebox{00}{\includegraphics{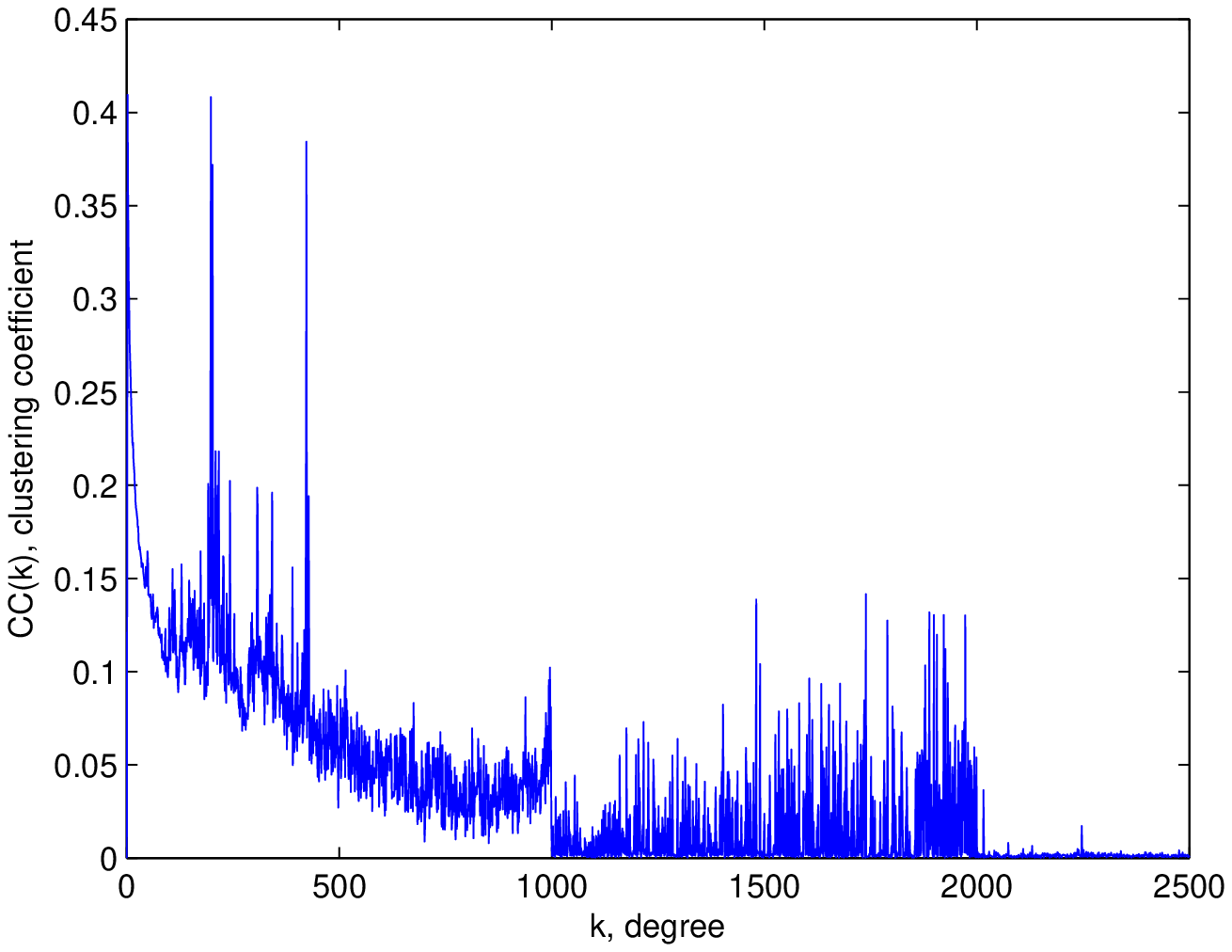}}}}
 %\vspace{0.3cm}
\caption{The clustering coefficient $C(k)$ against the node degree for LiveJournal. }
 \label{fig-3}
 \end{figure}

The LiveJournal system encourages users to bookmark and monitor particular blogs. This feature is exercised by virtually all users and results in a network of references between these blogs. The vast majority of blogs regularly read by a person are typically stored in the form of bookmarks as part of his profile. This degree of penetration of this behavior is driven by two main reasons. First, its convenience: it is impractical to periodically search for a particular blog and check whether it has new posts. The system automatically notifies users of the updates to the bookmarked blogs. Second, to protect their privacy, many users limit visibility of their posts to the users listed in their list of friends. Overall, the personal nature of these blogs and the intimate relationship between their authors give this network a powerful social aspect. In fact, we conducted a large number of case studies analyzing the threads of comments to verify that authors of many of the connected
  blogs actually know each other in person. It is therefore legitimate to refer to the network of blog bookmarks as social network. \\

In addition to personal profiles, users create communities that are in fact blogs run in collaboration by a number of users. Communities usually spin around a particular interest, well defined topic or represent a group of people united by a common task (such as, for instance, role-playing games) but otherwise are very similar to personal blogs. Periodic posts are discussed in threads of comments.  \\

LiveJournal has been used in a large number of academic studies \cite{2,3,4,5,6,7,8} due to its openness and availability of its well-designed APIs [http://www.livejournal.com/developer/]. In particular, all user profiles including the lists of monitored personal blogs and communities along with detailed information about the blog owners and their interests are freely accessible. \\

The data used in this work was obtained by crawling LiveJournal and collecting the entire content of all user profiles in the giant component. We defined the network nodes to correspond to personal blogs. Directional links connecting these nodes represent the record that a particular user (owning one blog) monitors another blog (owned by another user). We disregard community blogs as they usually do not represent individual users (but rather a group) and cannot be considered as part of the social network represented by personal blogs.\\

The social network obtained from the crawl contains 8.1 million users and over 125 million links. The average clustering coefficient is $C=0.1522$. However, having excluded nodes of degree 0 and 1, we get $C=0.2684$. The degree distribution is shown in Fig. \ref{fig-1}. The log-log plot reveals some deviations from linearity; we can distinguish two ranges of $k$, between 1 and 50 and between 50 and 1000, where it is linear. Still, the slopes
of the curves within these ranges do not differ much. The assortativity is confirmed by the data shown in Fig. \ref{fig-2}. There, the plots $<k'(k)>$
increase with $k$ both for in-going and out-going links.  \\

 \begin{figure}[ht]
 %\vspace{0.3cm}
 \centering
 {\centering \resizebox*{12cm}{9cm}{\rotatebox{-90}{\includegraphics{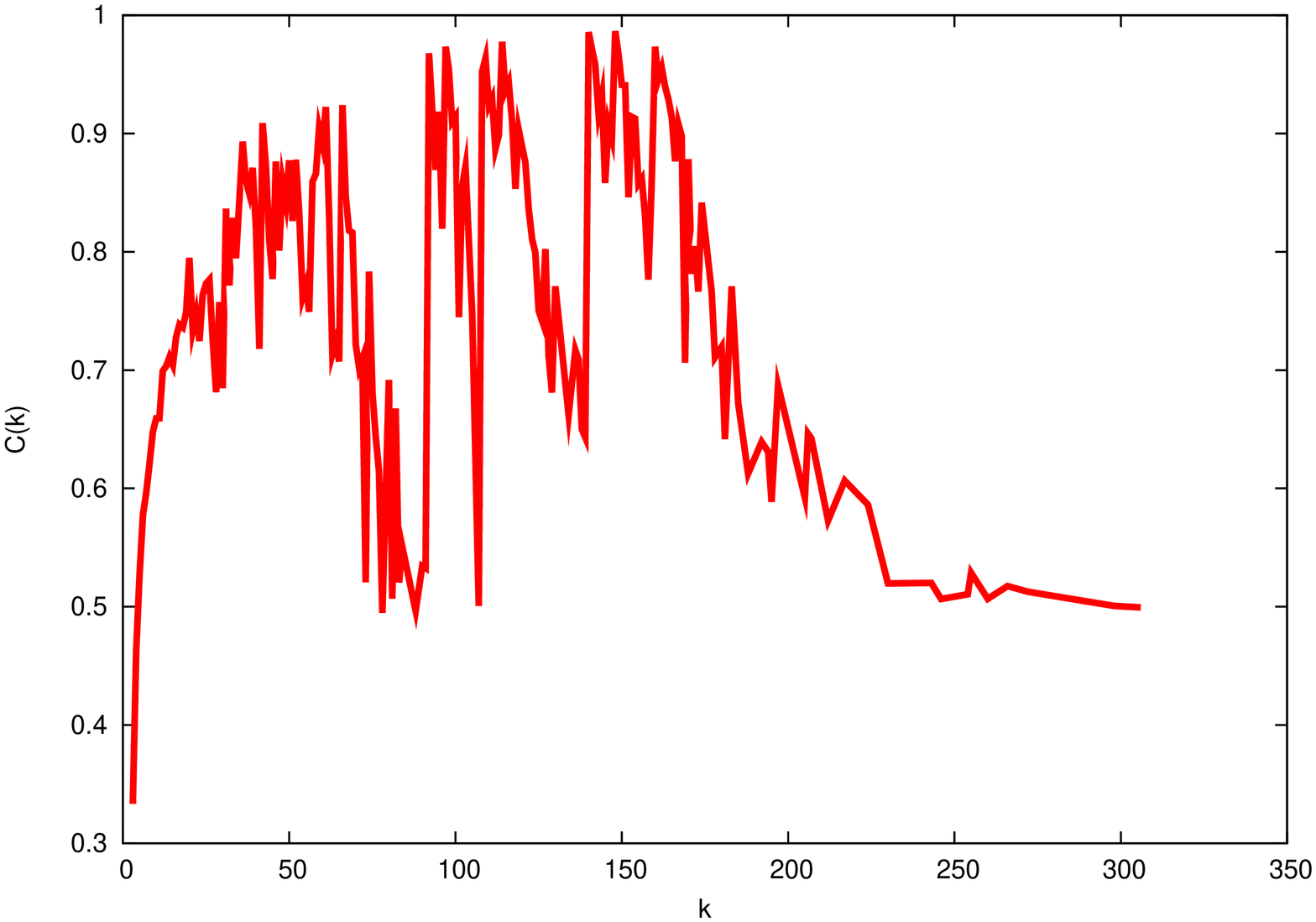}}}}
 %\vspace{0.3cm}
\caption{The clustering coefficient $C(k)$ for a line graph constructed on an uncorrelated scale-free network. }
 \label{fig-4}
 \end{figure}

As shown in Fig. \ref{fig-3}, the clustering coefficient $C(k)$ of LiveJournal varies strongly with $k$ in the social network. A strong local peak of the function $C(k)$ at given $k$ means that there is a lot of nodes of almost the same degree which are strongly clusterized. It is straightforward to imagine that these nodes belong to the same cluster - a clique. As cliques are formed from nodes in line graphs, the latter should display also the same kind of oscillations of the function $C(k)$. Indeed we found this behaviour for an artificially generated line graph of $10^4$ nodes, as shown in Fig. \ref{fig-4}. However we should note that the numerical values of the clustering coefficient $C$ are remarkably larger, than in the case of the data from LiveJournal. The reason for this discrepancy could lay in the fact that while the simulated line graph is uncorrelated,  the actual communities and interests grouping LiveJournal users are.

\section{Conclusion}

As we demonstrated above, the LiveJournal social network is scale-free, clustered and assortative. This makes it similar to the line graph, constructed on a scale-free network. Additionally, this similarity captures also the jagged character of the clustering coefficient dependence $C(k)$ on the node degree $k$. This similarity suggests, that a line graph, constructed on a scale-free network, is a fair representation of a realistic social network. This is the main goal of this paper. \\

Aside from suggesting a natural mechanism for the social network construction, a direct application of this result appears, if we are interested in a simulation of the process of spread of information, as alerts or gossips, in a community. For a large network, the direct simulation of the state of each particular node can be burdensome and memory consumming. Instead, we can consider a hypothesis that within the cliques, the information is shared almost immediately, when compared with the time of its transmission between the cliques. Such mechanism has been suggested by a number of social science researchers. In fact, it is implied by the Granovetter's groundbreaking "strength of the weak ties" \cite{gra}. If this is the case, it is possible to simulate the process on a much smaller network, where nodes represent cliques. \\

Concluding, topological arguments are presented that real social networks, where nodes represent agents, can be modeled as line graphs constructed on  initial networks, where nodes represent families, school classes, groups of friends who meet everyday, teams in firms etc. Modeling the spread of information, we can work on the initial networks, which are clearly smaller and more simple.

\section*{Acknowledgements} The research is partially supported within the FP7 project SOCIONICAL, No. 231288.

\end{document}